\documentclass[
  aps,             
  prb,               
  superscriptaddress,    
  twocolumn,             
  floatfix              
]{revtex4-2}

\usepackage{graphicx}     
\usepackage{txfonts}                
\usepackage{bm}

\usepackage{amsmath,amssymb}         
\usepackage{physics}               
\usepackage[colorlinks=true, linkcolor=blue, urlcolor=blue, citecolor=blue]{hyperref}

\usepackage{caption}                   

\begin{document}

\title{GPa Pressure Imaging Using Nanodiamond Quantum Sensors}

\author{Ryotaro Suda}
\email{	ryotaro.suda@phys.s.u-tokyo.ac.jp}
\affiliation{Department of Physics, The University of Tokyo, Bunkyo-ku, Tokyo 113-0033, Japan}

\author{Kenshin Uriu}
\affiliation{KYOKUGEN, Graduate School of Engineering Science, The University of Osaka, Toyonaka, Osaka 560-8531, Japan}

\author{Kouki Yamamoto}
\affiliation{Department of Physics, The University of Tokyo, Bunkyo-ku, Tokyo 113-0033, Japan}

\author{Misaki Sasaki}
\affiliation{Department of Physics, The University of Tokyo, Bunkyo-ku, Tokyo 113-0033, Japan}

\author{Kento Sasaki}
\affiliation{Department of Physics, The University of Tokyo, Bunkyo-ku, Tokyo 113-0033, Japan}

\author{Mari Einaga}
\affiliation{KYOKUGEN, Graduate School of Engineering Science, The University of Osaka, Toyonaka, Osaka 560-8531, Japan}

\author{Katsuya Shimizu}
\affiliation{KYOKUGEN, Graduate School of Engineering Science, The University of Osaka, Toyonaka, Osaka 560-8531, Japan}

\author{Kensuke Kobayashi}
\affiliation{Department of Physics, The University of Tokyo, Bunkyo-ku, Tokyo 113-0033, Japan}
\affiliation{Institute for Physics of Intelligence, The University of Tokyo, Bunkyo-ku, Tokyo 113-0033, Japan}
\affiliation{Trans-Scale Quantum Science Institute, The University of Tokyo, Bunkyo-ku, Tokyo 113-0033, Japan}

\date{\today}

\begin{abstract}
We demonstrate wide-field optical microscopy of the pressure distribution at \(\sim 20\)~GPa in a diamond anvil cell (DAC), using nitrogen-vacancy (NV) centers in nanodiamonds (NDs) as quantum sensors.  
Pressure and non-hydrostaticity maps are obtained by fitting optically detected magnetic resonance (ODMR) spectra with models incorporating hydrostatic and uniaxial stress conditions.  
Two methods for introducing NDs with a pressure-transmitting medium are compared, revealing that the embedding approach affects the degree of non-hydrostaticity. 
This ND-based technique offers a powerful imaging platform for probing pressure-induced phenomena and is extendable to other physical quantities such as magnetic fields.
\end{abstract}
\maketitle

\section{Introduction}
Pressure is a fundamental thermodynamic variable and has long been a key tuning parameter in condensed matter physics~\cite{Mao2018-sv}.  
Its application is of particular interest, as it can induce phase transitions and lead to emergent electronic properties.  
A notable example is oxygen, which becomes metallic and superconducting under high pressure~\cite{ShimizuNature1998,ShimizuJPSJ2005}, despite being a gas under ambient conditions.  
More recently, hydrogen-rich materials and nickelates have attracted significant attention as high-temperature superconductors under pressure~\cite{Drozdov2015-pm,Einaga2016-mv,DrozdovNature2019,SomayazuluPRL2019,SunNature2023}.

Progress in high-pressure research has been driven largely by advances in experimental techniques.
The diamond anvil cell (DAC) is now an essential tool, capable of generating pressures up to hundreds of gigapascals while allowing various in situ measurements~\cite{JayaramanRMP1983}.
Accurate pressure determination within the DAC sample chamber is crucial in such experiments.
Common methods include ruby fluorescence~\cite{Mao1976-kp,Mao1978-zu,Mao1986-kc}, diamond Raman spectroscopy~\cite{Hanfland1985-pz,Eremets2003-tq,Akahama2004-gj}, and diffraction-based techniques such as X-ray~\cite{Zha2000-ez} and neutron diffraction~\cite{Jorgensen1984-ik,Jorgensen1985-nq,Besson1992-xn}.
Each approach offers unique advantages depending on the experimental context.

Quantum sensing based on nitrogen-vacancy (NV) centers in diamonds has recently emerged as a powerful technique for local measurement of physical quantities~\cite{Taylor2008, Balasubramanian2008, Rondin2014,Schirhagl2014-dw,DegenRMP2017,Casola2018}. 
Their electron spin resonance can be optically detected via photoluminescence (PL) under simultaneous laser and microwave irradiation, a technique known as optically detected magnetic resonance (ODMR).  

This enables non-invasive measurement of magnetic fields, temperature, and pressure.  
Embedding NV centers near the culet surface of a DAC facilitates probing of pressure and magnetic fields in close proximity to the sample, which has been difficult using conventional techniques.  
This approach has recently been employed to study novel phenomena such as magnetic phase transitions in iron and superconducting transitions in hydrides~\cite{Steele2017-po,Hsieh2019-op,Lesik2019-vw,Ho2020-cb,Shelton2024-kl,Wang2024-rf,Yip2019-of}.

In this context, several studies have proposed an alternative approach in which microdiamonds~\cite{Yip2019-of,Shang2019-ol} or nanodiamonds (NDs)~\cite{Shelton2024-kl,Ho2023-dp} containing NV centers are embedded directly into the DAC sample chamber.  
Unlike the method of embedding NV centers into the culet, this method places NV centers within the chamber, allowing more direct probing of the sample's local environment.  
As hard diamond particles are introduced into the chamber with pressure-transmitting medium (PTM), it is crucial to identify embedding strategies compatible with the desired pressure conditions.  
Moreover, the use of ND ensembles introduces crystal orientation variations that complicate ODMR spectra, highlighting the need for analytical techniques suitable for optical imaging with CMOS cameras.

In this study, we perform micrometer-scale imaging of the pressure distribution inside a DAC sample chamber using NV centers in a layered ensemble of NDs.
By analyzing ODMR spectra with models assuming hydrostatic or uniaxial stress environments, we extract the pressure along the loading $P_Z$ and transverse $P_\perp$ directions, as well as the anisotropy ratio $\lambda = P_Z / P_\perp$ to characterize non-hydrostaticity.
We further investigate methods for introducing ND ensembles in the sample chamber and experimentally demonstrate that the arrangement of the sample, PTM, and NDs significantly affects the degree of non-hydrostaticity.
Our method, capable of operating at pressures as high as 20~GPa, is anticipated to be applicable to property measurements of a wide variety of materials under high pressure.

This paper is organized as follows. Section 2 presents the principles of quantum sensing using NV centers. Section 3 describes the experimental setup, including the configuration of the sample and the PTM inside the DAC chamber. The procedure for acquiring ODMR spectra is also detailed. In Section 4, we present imaging results of the pressure distribution in the chamber based on the model described in Sec.~2, and evaluate the non-hydrostaticity in the chamber. Finally, Section 5 summarizes our findings with future prospects.

\section{Principles}
\subsection{NV Centers}

\begin{figure}
  \begin{center}
    \includegraphics[width=\linewidth]{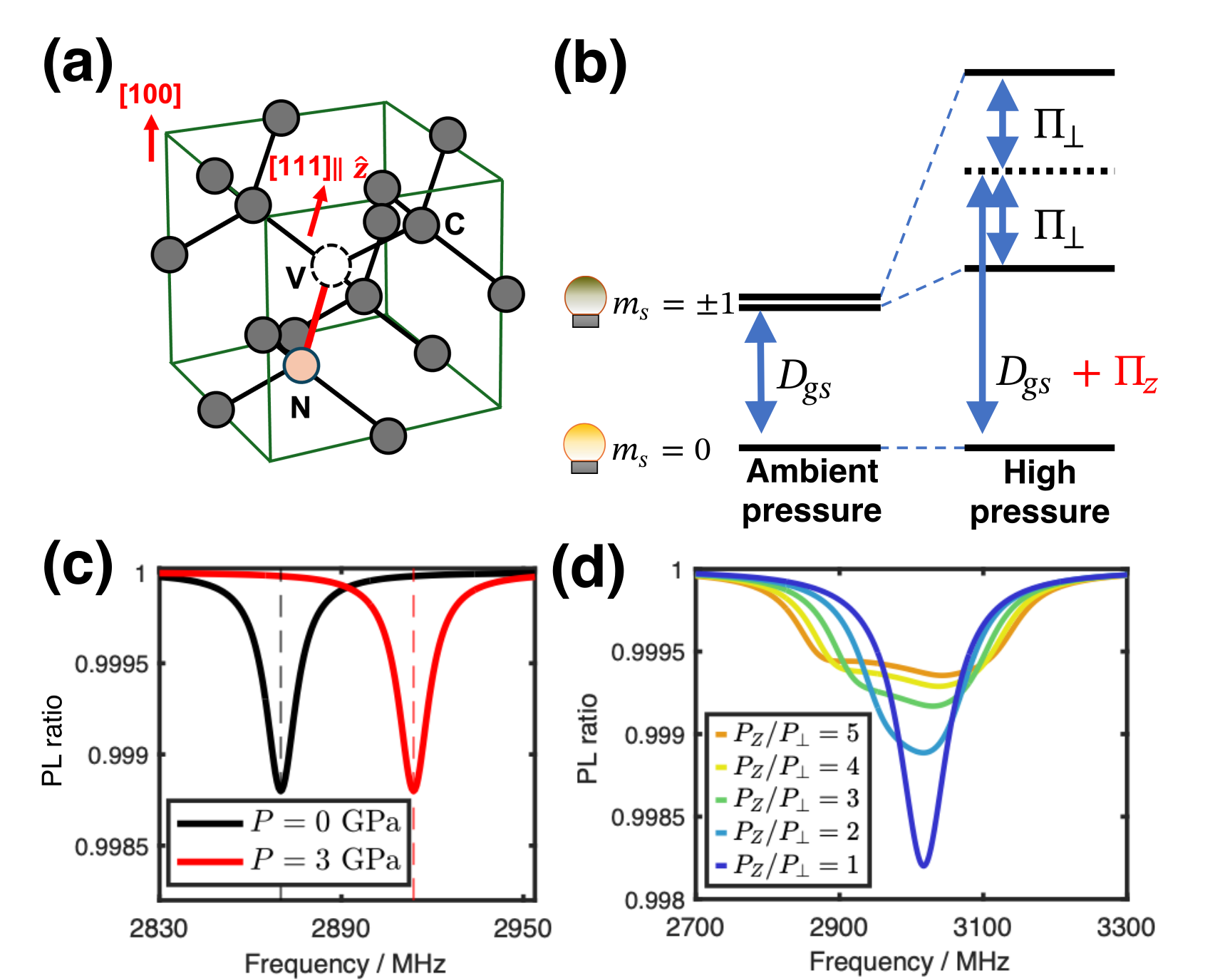}
  \end{center}
\caption{\label{fig:1} (Color online) 
(a) Structure of an nitrogen-vacancy (NV) center. It consists of a nitrogen (N) atom and a vacancy (V) substituting two adjacent carbon (C) atoms in the diamond lattice. The local coordinate system is defined such that the [111] direction corresponds to $\hat{z}$. 
(b) Energy-level diagram of the ground state of an NV center and its shift under pressure. 
(c) Simulated ODMR spectra ($I_{\mathrm{NV}}$) under hydrostatic pressure (ODMR amplitude \(C=0.0006\), linewidth \(d\nu=12\ \mathrm{MHz}\)). 
(d) Simulated ODMR spectra ($I_{\mathrm{NDs}}$) under uniaxial stress (ODMR amplitude \(C=0.02\), linewidth \(d\nu=40\ \mathrm{MHz}\)). 
The PL ratio, shown on the vertical axes of (c) and (d), is defined as the ratio of the photoluminescence (PL) intensity with and without microwave irradiation.
}
\end{figure}

An NV center consists of a substitutional nitrogen atom adjacent to a vacancy, as illustrated in Fig.~\ref{fig:1}(a).  
This defect has \(C_{3v}\) symmetry, with its axis, known as the NV axis, aligned along one of the \(\langle111\rangle\) crystallographic directions.  
The negatively charged NV center has an electron spin of \(S = 1\)~\cite{Loubser1978-lc}, and we refer to this charge state simply as the NV center in the following.
The spin energy levels in the electronic ground state are shown in Fig.~\ref{fig:1}(b).  
The spin states with \(m_S = 0\), \(-1\), and \(+1\) correspond to projections of spin angular momentum along the NV axis.
In the absence of a magnetic field and at ambient pressure, the \(m_S = 0\) and \(m_S = \pm 1\) states are split by the zero-field splitting \(D_{\mathrm{gs}} = 2.87~\mathrm{GHz}\).~\cite{Schirhagl2014-dw}

External perturbations such as stress, magnetic field, or temperature can shift these energy levels.  
In particular, stress modifies the electron distribution and symmetry of the defect, resulting in mixing and shifts of the spin states.  
Further details are provided in Sec.~\ref{sec2.2}.

The spin states of the NV center can be polarized or initialized to the \(m_S = 0\) state via optical excitation and read out through PL, owing to spin-dependent optical transitions.  
This enables ODMR spectroscopy, where electron spin resonance is measured by monitoring PL intensity under continuous laser illumination while sweeping the microwave frequency.  
Figure~\ref{fig:1}(c) shows simulated ODMR spectra. The vertical axis indicates the PL ratio, which is defined as the PL intensity with microwave irradiation divided by that without. A dip appears when the microwave frequency matches the spin resonance.
At zero magnetic field and ambient pressure, a dip appears at the zero-field splitting \(D_{\mathrm{gs}} = 2.87~\mathrm{GHz}\).  
Shifts in energy levels lead to changes in the shape and position of the ODMR spectrum.  
Thus, physical quantities such as pressure, magnetic field, and temperature can be extracted from ODMR measurements~\cite{Balasubramanian2008,Hsieh2019-op,Maze2008-or,Acosta2010-lt,Barfuss2019-db,Yamamoto2025-df}. 
This quantum sensing technique, based on optical detection, is compatible with wide-field microscopy for imaging applications.~\cite{Pham2011-bk}

In this study, we use NV centers in ND ensembles embedded in the DAC sample chamber.  
This approach enables micrometer-scale spatial resolution via wide-field microscopy while preserving the localized sensitivity of the NV centers.  
Using this method, we evaluate the pressure distribution in the DAC sample chamber as a function of the spatial arrangement of ND ensembles and PTM.

\subsection{Pressure Sensing via ODMR Spectra}\label{sec2.2}

In this subsection, we describe the stress dependence of an individual NV center in a ND.  
The ODMR spectra of NV centers in ND ensembles are discussed in the following subsection.  
To model the stress dependence, we consider the effective spin Hamiltonian~\cite{Steele2017-po,Hsieh2019-op,Barfuss2019-db,Doherty2014-gg,Doherty2012-mi,Barson2017-zy,Udvarhelyi2018-if,felton2009hyperfine}.  
In this study, we neglect external perturbations such as magnetic fields and employ the following spin Hamiltonian:
\begin{align}\label{hamiltonian}
    \hat{H}/h = &(D+\Pi_z) \hat{S}_z^2 +\Pi_x (\hat{S}_x^2 - \hat{S}_y^2) + \Pi_y (\hat{S}_x \hat{S}_y + \hat{S}_y \hat{S}_x),
\end{align}
where \(\hat{S}_{x,y,z}\) are spin-1 operators along the \(x\), \(y\), and \(z\) directions.  
The NV coordinate system is defined as \(x\): [1 1 \(\bar{2}\)], \(y\): [1 \(\bar{1}\) 0], and \(z\): [1 1 1], with the \(z\)-axis aligned with the NV axis [see Fig.~\ref{fig:1}(a)].  
The stress-induced energy shifts \(\Pi_{x,y,z}\) are given by [see Fig.~\ref{fig:1}(b)]:
\begin{align}
\Pi_x &= \alpha_2(\sigma_{yy} - \sigma_{xx}) + \beta_2(2\sigma_{xz}),\label{eq2} \\ 
\Pi_y &= \alpha_2(2\sigma_{xy}) + \beta_2(2\sigma_{yz}),\label{eq3} \\ 
\Pi_z &= \alpha_1(\sigma_{xx} + \sigma_{yy}) + \beta_1\sigma_{zz},\label{eq4}
\end{align}
where \(\sigma_{ij}\) are components of the local stress tensor. 
The coefficients are known as \(\alpha_1 = 8.6\)~MHz/GPa, \(\alpha_2 = -1.95\)~MHz/GPa, \(\beta_1 = -2.5\)~MHz/GPa, and \(\beta_2 = -4.5\)~MHz/GPa~\cite{Hsieh2019-op,Shelton2024-kl}.  
The resonance frequencies \(f_{\pm}\), obtained by diagonalizing the Hamiltonian, are
\begin{align}\label{para_energy_2}
    f_{\pm} = D_{\mathrm{gs}} + \Pi_z \pm \Pi_\perp,
\end{align}
where \(\Pi_\perp = \sqrt{\Pi_x^2 + \Pi_y^2}\).  
These resonance frequencies determine the position and shape of the ODMR spectrum.  
The spectrum is described by a sum of Lorentzian functions centered at \(f_{\pm}\):
\begin{align}
    I_{\mathrm{NV}} = 1 - [L(f_{\mathrm{mw}},f_{-},\nu,C) + L(f_{\mathrm{mw}},f_{+},\nu,C)],\label{eq6}
\end{align}
where \(L(f_{\mathrm{mw}}, f_{\mathrm{res}}, \nu, C) = \frac{C}{1 + \left(\frac{f_{\mathrm{mw}} - f_{\mathrm{res}}}{\nu}\right)^2}\) is a Lorentzian function centered at \(f_{\mathrm{res}}\) with amplitude \(C\) and linewidth \(\nu\).

Then, we describe the ODMR spectrum under hydrostatic pressure, the simplest pressure condition.
Under hydrostatic conditions (i.e., \(\sigma_{xx} = \sigma_{yy} = \sigma_{zz}\) and \(\sigma_{ij} = 0\) for \(i \neq j\)), the stress tensor reduces to a scalar matrix, with its diagonal elements corresponding to the pressure \( P \).
Figure~\ref{fig:1}(c) presents simulated spectra at ambient pressure (\( P = 0~\mathrm{GPa} \)) and under an applied pressure of \( P = 3~\mathrm{GPa} \), shown by the black and red lines, respectively.
The spectral shape remains unchanged, whereas the dip position, i.e., the resonance frequency, shifts with increasing pressure.
Note that  under hydrostatic conditions the transverse parameter $\Pi_\perp = 0$ [see  Eqs.~(\ref{eq2}) and (\ref{eq3})] with only  $\Pi_z$  finite.
The parameter \(\Pi_z\) directly corresponds to the shift in the resonance frequency and increases linearly with \( P \).
\(\Pi_z\) is calculated as \(\Pi_z = (2\alpha_1 + \beta_1)P = 14.7~\mathrm{MHz/GPa} \times P = 44.1~\mathrm{MHz}\) for \( P = 3~\mathrm{GPa} \), corresponding to the difference in dip position between the black and red spectra in Fig.~\ref{fig:1}(c).
Since the effective spin Hamiltonian provides the relationship between the parameters \(\Pi_z\) and \(\Pi_\perp\) and the stress tensor, the pressure environment can be probed from the ODMR spectrum.

\subsection{ODMR Spectrum under Uniaxial Stress}

\begin{figure}
\centering
\includegraphics[width=\linewidth]{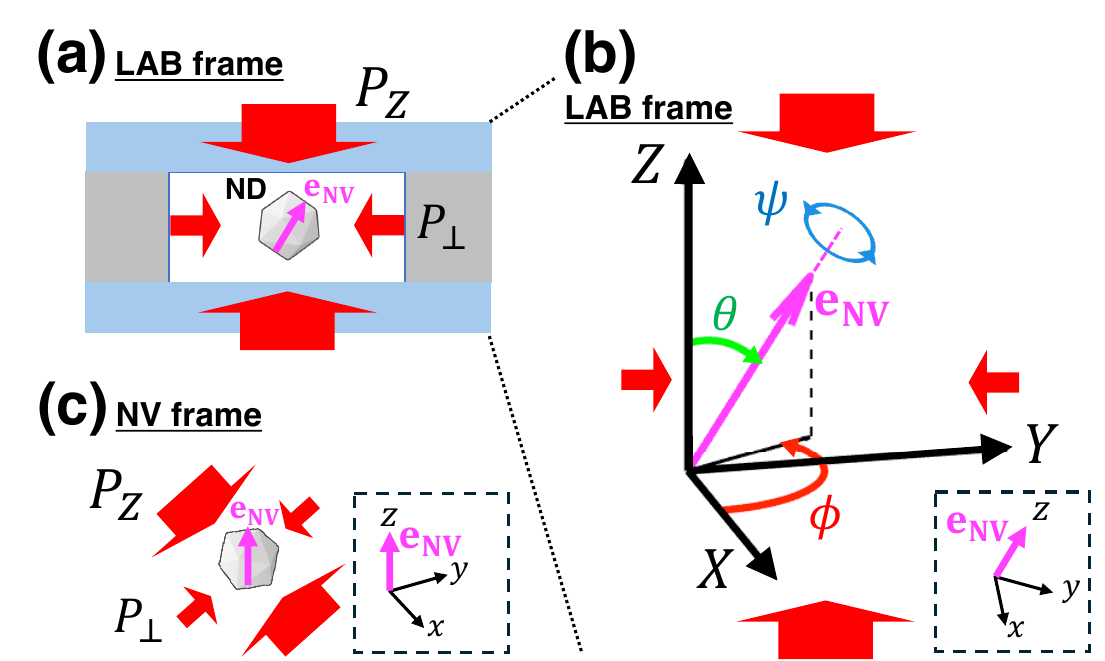}
\caption{\label{fig:6} (Color online) 
(a) Schematic of a ND under uniaxial stress in the DAC sample chamber. The vertical and horizontal components correspond to $P_Z$ and $P_\perp$, respectively, in the LAB frame.  
(b) Orientation of the NV axis ($\bm{e}_{\mathrm{NV}}$) in the LAB frame, represented using the polar angle $\theta$, azimuthal angle $\phi$, and crystal rotation angle $\psi$.    
(c) Schematic of a ND under uniaxial stress in the NV frame, where the NV axis is defined as the $z$-axis.
}
\end{figure}

The stress environment inside the DAC sample chamber is often not purely hydrostatic.
Instead, as illustrated in Fig.~\ref{fig:6}(a), a uniaxial stress condition frequently develops, characterized by a higher pressure along the loading axis (\( P_Z \)) than in the transverse (\( P_{\perp} \)) directions~\cite{Hsieh2019-op,Shelton2024-kl,Wang2024-rf,Ho2023-dp,Bhattacharyya2024-oa}. 
In the laboratory coordinate system $(X, Y, Z)$, where the DAC loading axis is defined as the \( Z \)-axis, such uniaxial stress environment is expressed as follows:
\begin{align}
\sigma_{\mathrm{LAB}} = 
\begin{pmatrix} P_{\perp} & 0 & 0 \\ 0 & P_{\perp} & 0 \\ 0 & 0 & P_Z \end{pmatrix},
\end{align}
where \(P_Z\) and  \(P_\perp\) are axial and transverse pressure, respectively.

When NDs containing NV centers are incorporated into a DAC sample chamber, their orientations become random [Fig.~\ref{fig:6}(a)].
As a result, stress components that break the symmetry of the NV centers and split the ODMR spectra, which could be neglected under hydrostatic conditions, can no longer be ignored.
Even when the uniaxial stress happens to align with the NV axis, the resonance frequency varies depending on the magnitude of the uniaxial stress, characterized by \(\lambda = P_{Z}/P_{\perp}\).
Furthermore, considering that the NDs exist as an ensemble, the overall spectral shape can be estimated by summing the signals from NV centers with various orientations.
Although the situation becomes complex compared to the hydrostatic case, the ODMR spectrum can still be reproduced by combining existing models such as the effective spin Hamiltonian.
By utilizing this approach, the stress components can be estimated.
In the following, we describe the step-by-step process for calculating the ODMR spectra of NV centers in ND ensembles embedded in the DAC sample chamber.
First, we select a specific NV center and describe how the uniaxial stress is incorporated into its spin Hamiltonian.
Then, by summing the PL signals from NV centers with various orientations, we formulate the ODMR spectrum that would be obtained in the experiments conducted in this study.

We consider an NV center whose NV axis is oriented at a polar angle \(\theta\) and an azimuthal angle \(\phi\) with respect to the laboratory coordinate system [see Fig.~\ref{fig:6}(b)].
The relationship between the NV coordinate system \((x, y, z)\), as defined in the previous subsection, and the laboratory coordinate system \((X, Y, Z)\) is given as follows:
\begin{align}
\mathbf{e}_{x} &= U \mathbf{e}_X,  \\
\mathbf{e}_{y} &= U \mathbf{e}_Y, \\
\mathbf{e}_{z} &= \mathbf{e}_{\mathrm{NV}} = U \mathbf{e}_Z, \\
U &= R_{\mathrm{NV}}(\psi) R_Z(\phi) R_Y(\theta),
\end{align}
where \(\mathbf{e}_{k}\) denotes the unit vector along the \(k\)-axis, and \(R_k(\omega)\) represents the rotation matrix corresponding to a counterclockwise rotation by an angle \(\omega\) about the \(k\)-axis.
Here, the rotation matrix \(R_{\mathrm{NV}}(\psi)\) is required to rigorously account for the local \((x, y, z)\) coordinate system of the NV center and its crystal structure [see left bottom axes in Fig.~\ref{fig:6}(b)].
By using the transformation matrix \(U\) between the coordinate systems, the stress tensor $\sigma$ in the NV coordinate system [Fig.~\ref{fig:6}(c)] can be obtained as follows:
\begin{align}
\sigma = U\, \sigma_{\mathrm{LAB}}\, U^T.
\end{align}
By calculating the parameters \(\Pi_x\), \(\Pi_y\), and \(\Pi_z\) from the components of the stress tensor in the NV coordinate system obtained above, according to Eqs.~(\ref{eq2})--(\ref{eq4}), the resonance frequencies and ODMR spectrum for the specific NV center can be determined using Eq.~(\ref{para_energy_2}) and Eq.~(\ref{eq6}), respectively.

To estimate the ODMR spectrum of ND ensemble measurements, the signals from all possible NV center orientations \((\theta, \phi, \psi)\) must be summed.
This summation must account for both variations in the ODMR spectra and differences in PL intensity, arising from the anisotropy of collection efficiency $A$ and absorption efficiency $\kappa$\cite{Horowitz2012-nw,Tsukamoto2022-sb}.

Applying the appropriate weighting yields the following expression:
\begin{align}\label{equni1}
I_{\mathrm{NDs}} \;=\;&
1 - \int_{0}^{2\pi} \!d\psi
\int_{D} d\phi\,d\theta\,\sin\theta\, 
[(1-I_{\mathrm{NV}}) \nonumber\\
&\times M(\theta,\phi)\,\kappa(\theta,\phi)\,A(\theta,\theta_{\text{max}})]+\delta,
\end{align}
where $I_{\mathrm{NDs}}$ is the ODMR spectrum of the ND ensemble measurement, $I_{\mathrm{NV}}$ is the ODMR spectrum shape from the NV center at a specific orientation $(\theta,\phi,\psi)$, \(D\) denotes the integration range determined by the numerical aperture (NA) of the objective lens, where \(\mathrm{NA}/ n_d  = \sin(\theta_{\mathrm{max}})\) with $n_d$ the refractive index of diamond\cite{thomas1993optical}, and $\delta$ denotes an offset term.

The other terms in the Eq.~(\ref{equni1}) are defined as follows:
\begin{align}
A(\theta,\theta_{\text{max}}) &\propto \frac{\pi}{12} 
\left[ 32 - \cos\theta_{\text{max}} \left(31 + \cos(2\theta_{\text{max}}) \right. \right. \nonumber \\
&\quad \left. \left. - 6\cos(2\theta)\sin^2\theta_{\text{max}} \right) \right], \label{equni10}
\\
\kappa(\theta)&\propto 1+\cos^2{\theta}, \label{equni11}
\\
M(\theta,\phi) &=(1-\cos\phi\sin\theta)^2. \label{equniS}
\end{align}
To account for not only the optical measurement anisotropy but also the directional dependence of the microwave magnetic field, we introduce a weighting factor \(M\).
The ODMR contrast of an NV center increases with the amplitude of the microwave magnetic field component perpendicular to the NV axis.
The factor \(M\) is based on the assumption that, when the microwave amplitude is weak, the contrast depends quadratically on the microwave amplitude.~\cite{Shelton2024-kl,Dreau2011-gc,Jensen2013-nb}

Note that we assumed the ODMR amplitude \(C\) and linewidth \(d\nu\) to be identical for all NV centers.

Figure~\ref{fig:1}(d) shows the simulated ODMR spectrum of ND ensemble under uniaxial stress. 
In this simulation, the total pressure was fixed as $P_Z + 2P_\perp = 30~\mathrm{GPa}$. As the uniaxiality $\lambda = P_Z / P_\perp$ increases, the overall spectral linewidth broadens into a trapezoidal, non-Lorentzian shape, the spectrum becomes asymmetric, and the contrast is reduced. By fitting the experimentally obtained ODMR signal with this model, the pressures can be estimated.

Under the assumption of a uniaxial stress environment, the fitting parameters are only five: \(C\), \(d\nu\), \(P_Z\), \(P_{\perp}\), and $\delta$.
The small number of fitting parameters makes it suitable for imaging measurements that require fitting a large number of ODMR spectra.
Under hydrostatic conditions, \(P = P_Z = P_\perp\), which reduces the number of fitting parameters and simplifies the spectral shape to the form given by Eq.~(\ref{eq6}).

A similar model has been proposed by Shelton \emph{et al.}~\cite{Shelton2024-kl}, but our approach subsumes both hyperfine and electric field broadening into a single Lorentzian linewidth \(d\nu\), while also incorporating the anisotropy of collection efficiency \(A(\theta,\theta_{\max})\) and absorption efficiency \(\kappa(\theta,\phi)\)~\cite{Horowitz2012-nw,Tsukamoto2022-sb}. This description retains sufficient information for ODMR spectrum analysis and makes our model particularly well suited for rapid, large-area pressure imaging measurements.  

\section{Experiment}
\subsection{Experimental setup}
\begin{figure}
\centering
\includegraphics[width=\linewidth]{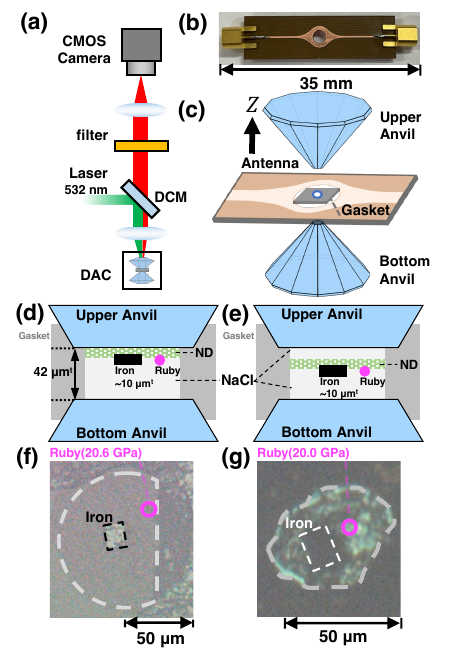}
\caption{\label{fig:2} (Color online) 
(a) A custom wide-field microscopy system used in this study.  
(b) A custom designed microstrip-line-based antenna with an aperture. 
The antenna is fabricated by UV laser machining from one side of a PCB substrate composed of a \(0.3\,\mathrm{mm}\)-thick FR4 with \(18\,\mu\mathrm{m}\)-thick copper foil on both sides.  
This structure enables broadband and spatially uniform microwave magnetic field irradiation into the sample chamber, which is located slightly above the central hole, with the field oriented perpendicular to the direction of the microstrip line.
(c) Schematic of the DAC. 
Pressure is applied by sandwiching a rhenium gasket with a sample hole between the upper and bottom diamond anvils. 
Microwaves are applied to the DAC sample chamber using the antenna shown in (b). 
The laboratory coordinate system is taken so that the vertical compression direction corresponds to the $Z$-axis.
A small gap is introduced between the gasket and the antenna to prevent degradation of the antenna's performance.
(d, e) Cross-sectional schematics of the DAC sample chamber in the single-layer PTM configuration (d) and in the double-layer PTM configuration (e).  
(f, g) Microscope images of the DAC sample chamber taken from above. The gray outline indicates the sample chamber and the pink circle marks the ruby used for pressure calibration. For future magnetic field evaluation, $4\mathrm{N}$ purity iron is placed in the dashed box region. (f) corresponds to the single-layer PTM configuration, and (g) to the double-layer PTM configuration.}
\end{figure}

We develop a custom wide-field microscopy system to image stress distributions via ODMR spectroscopy using NDs embedded in the DAC sample chamber [Fig.~\ref{fig:2}(a)].  
A continuous-wave green laser (\(532\,\mathrm{nm}\)) excites the NV centers through an objective lens.  
The resulting red fluorescence is collected by the same lens, transmitted through a dichroic mirror (DCM) and optical filters, and imaged onto a CMOS camera via an achromatic lens.  
To maintain fluorescence contrast, a \(532\,\mathrm{nm}\) notch filter, a \(650\,\mathrm{nm}\) long-pass filter, and an \(800\,\mathrm{nm}\) short-pass filter are employed.  
Microwaves are applied to the DAC sample chamber via a custom designed microstrip-line-based antenna with an aperture [Fig.~\ref{fig:2}(b)]. 
Each image pixel corresponds to NV centers at a specific location within the sample chamber, allowing spatially resolved stress mapping by analyzing the ODMR spectra pixel by pixel.  

The objective and imaging lenses are adjusted for each measurement.  
All experiments are conducted at room temperature without an external magnetic field.

The pressure is applied by sandwiching the sample between the upper and bottom diamond anvils [Fig.~\ref{fig:2}(c)].
We use a pair of diamond anvils with a culet diameter of $260\,\mathrm{\mu m}$ and a rhenium (Re) gasket with an initial thickness of $250\,\mathrm{\mu m}$. It is pre-indented to a thickness of $42\,\mathrm{\mu m}$ using the diamond anvils, and a $100\,\mathrm{\mu m}$ diameter hole is drilled at the center using an infrared laser to form the sample chamber.
The DAC sample chamber is filled with sodium chloride ($\mathrm{NaCl}$) as a PTM, along with 50~nm- sized NDs (Ad\'{a}mas Nanotechnologies, NDNV50nmHi10ml) for measurement and a ruby crystal of about $10~\mu$m in size for pressure calibration. 
A small iron sample (4N purity, Nilaco Co.; $\sim 10\,\mu\mathrm{m}^t$) is also embedded for future magnetic field evaluation.

\subsection{Arrangement of PTM}

To investigate the methods for arranging NDs inside the sample chamber, we conduct two types of experiments by modifying the arrangement of the NDs and PTM within the sample chamber.

\subsubsection{Single-layer PTM}
The first configuration models the situation in which the sample (NDs, ruby, and iron) is pressed directly against the culet surface of the upper diamond anvil without any surrounding PTM.

In this configuration, NaCl is placed only below the ND layer in the sample chamber [Figs.~\ref{fig:2}(d) and (f)]. First, NaCl powder is loaded at the bottom of the chamber as the PTM, followed by the placement of the ruby crystal and the iron sample on top of it, and finally the ND layer is placed above them.

Figure~\ref{fig:2}(d) shows a schematic cross-sectional view of the sample chamber. Figure~\ref{fig:2}(f) is a microscope image taken from the top of the chamber. The gray dashed line indicates the boundary of the sample chamber, and the pink circle marks the location of the ruby used for pressure calibration. 
The pressure obtained from the ruby fluorescence is $20.6 ~\mathrm{GPa}$~\cite{Mao1978-zu}.  In this configuration, the NDs, the ruby, and iron sample are in contact with the upper diamond anvil.

\subsubsection{Double-layer PTM}
The second configuration models the situation in which the sample is fully surrounded by the PTM, preventing direct contact with the diamond culet.

In this configuration, NaCl is placed both below and above the ND layer in the sample chamber, so that the ND layer is sandwiched between two layers of PTM [Figs.~\ref{fig:2}(e) and (g)]. First, NaCl powder is loaded at the bottom of the chamber, followed by the placement of the ruby and iron on top of it, and then the ND layer is placed above them. Finally, an additional layer of NaCl powder is set on top of the ND layer to complete the double-layer PTM configuration.

Figure~\ref{fig:2}(e) shows a schematic cross section of the chamber, and Fig.~\ref{fig:2}(g) is a microscope image taken from the top. As in the previous case, the gray dashed line indicates the sample chamber, and the pink circle indicates the ruby position. 

Note that both the NDs and NaCl are too small and transparent to be visible in the optical microscope images.
The pressure obtained from the ruby fluorescence is $20.0~\mathrm{GPa}$~\cite{Mao1978-zu}. In this configuration, an additional NaCl layer is placed above the ND layer and measurement samples, so that the NDs, ruby, and iron are not in contact with the culet surface of the upper diamond anvil.

\section{Results and Discussion}
\subsection{Single-layer PTM Configuration}

\begin{figure}
\centering
\includegraphics[width=\linewidth]{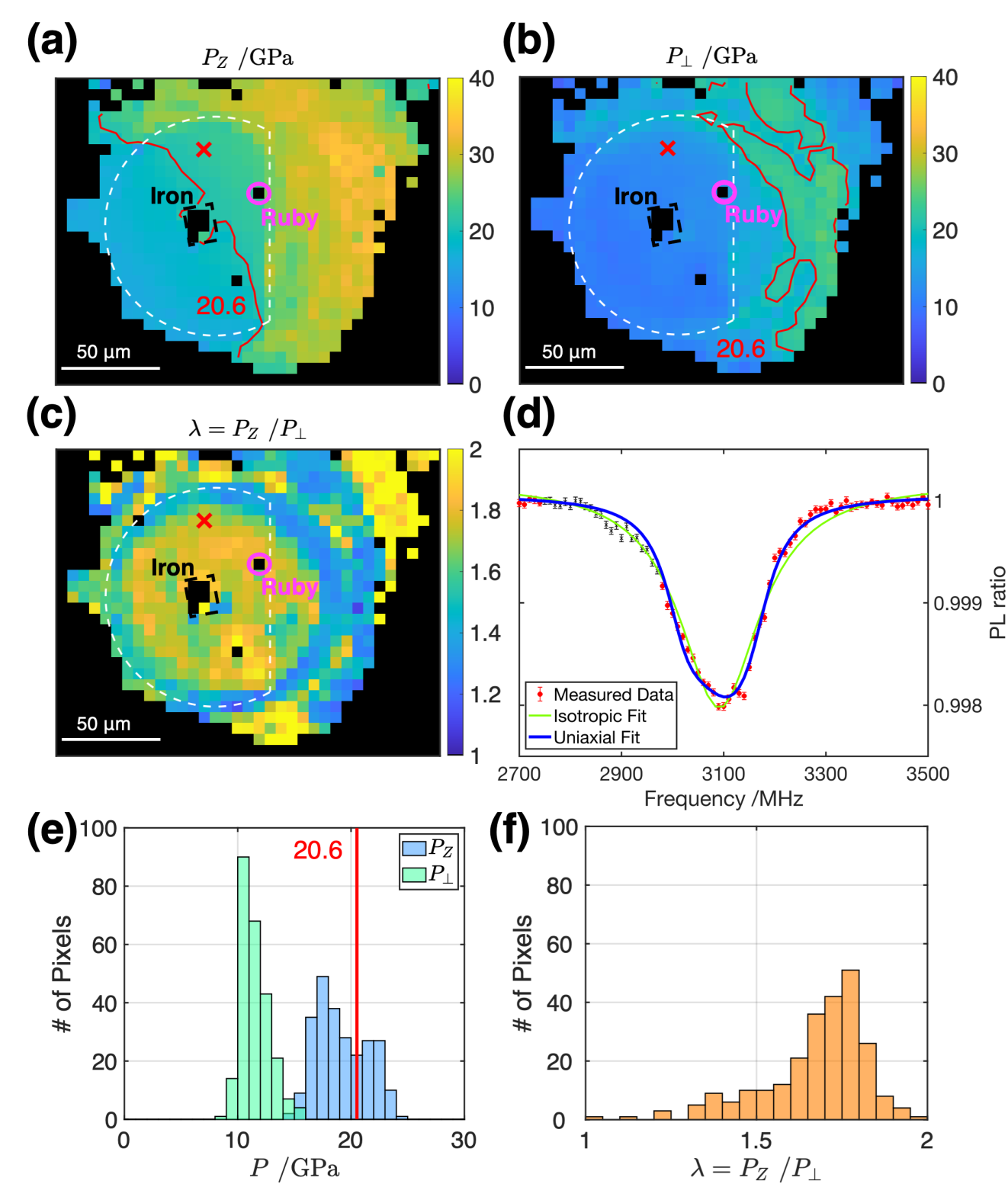}
\caption{\label{fig:3} (Color online) 
Results from the single-layer PTM experiment. 
(a, b, c) Spatial distributions of (a) the axial pressure $P_Z$, (b) the transverse pressure $P_\perp$, and (c) the pressure anisotropy ratio $\lambda = P_Z / P_\perp$. The white dashed line indicates the boundary of the sample chamber, the pink circle marks the location of the ruby, the black box indicates the iron sample. The black pixels denote regions with insufficient signal intensity. In (a) and (b), the red solid line corresponds to the pressure value of $20.6\,\mathrm{GPa}$ obtained from the ruby.  
(d) ODMR spectrum at the location marked by a red cross in (a)--(c). Black points are excluded from the fitting to avoid the influence of ambient pressure components.
(e) Histogram of $P_Z$ and $P_\perp$ within the sample chamber.  
(f) Histogram of the pressure anisotropy ratio $\lambda = P_Z / P_\perp$ within the sample chamber.
}
\end{figure}

In the single-layer PTM configuration, clear signatures of uniaxial stress are evident from the spatial pressure distributions in the sample chamber, as shown in Figs.~\ref{fig:3}(a) and (b). Before discussing these images, we analyze a representative ODMR spectrum presented in Fig.~\ref{fig:3}(d). This spectrum is obtained at the location marked by a red cross in Figs.~\ref{fig:3}(a) and (b). The experimental spectrum, shown in red, exhibits a central frequency of approximately 3100~MHz and a linewidth of about 200~MHz, featuring an asymmetric dip shape. This central frequency corresponds to approximately 15~GPa, assuming a purely hydrostatic pressure condition. 

Figure~\ref{fig:3}(d) indicates a slight dip around 2900~MHz, which corresponds to the resonance frequency of NV centers at ambient pressure. This feature arises from signals originating from NDs that are not under pressure and have unintentionally contributed to the spectrum. Data in this frequency range (indicated by black markers) are excluded from the following analysis.

This spectrum is fitted using the hydrostatic model (green line) and the uniaxial stress model (blue line)~[Fig.~\ref{fig:S1}]. The residuals in the high-frequency region (3300--3450~MHz) are significantly larger for the hydrostatic model, whereas the uniaxial model more accurately reproduces the observed asymmetry. From the fit, we obtain $P_Z=22.1$~GPa and $P_\perp=12.5$~GPa, yielding a pressure anisotropy ratio $\lambda = P_Z / P_\perp=1.77$.

Based on these results, we employ a uniaxial stress model to extract the axial pressure (\(P_Z\)) and transverse pressure (\(P_\perp\)).  
Their spatial distributions are shown in Figs.~\ref{fig:3}(a) and \ref{fig:3}(b), respectively.  
A histogram of the pressure distribution within the chamber is presented in Fig.~\ref{fig:3}(e), where the ruby-measured pressure (20.6~GPa) is indicated by a vertical red line.  
The axial pressure \(P_Z\) ranges from 14 to 30~GPa, while the transverse pressure \(P_\perp\) varies from 10 to 20~GPa and consistently remains lower than \(P_Z\).

Although accurate pressure determination by ruby fluorescence is technically difficult under non-hydrostatic conditions,~\cite{Piermarini1973-jb,Fujishiro1988-lc,He1995-nt} the pressure estimated from NV centers agrees, at least in order of magnitude, with the ruby result, supporting the validity of the present model.

The distribution of the anisotropy ratio \(\lambda = P_Z / P_\perp\) and its histogram are shown in Figs.~\ref{fig:3}(c) and \ref{fig:3}(f), respectively.  
In the central region of the sample chamber, \(\lambda\) ranges from approximately 1.6 to 1.8, while it decreases to around 1.4 near the chamber boundaries.  

This trend suggests that gasket-induced transverse pressure reduces uniaxial stress near the edges of the chamber.  
Although some variation exists, \(\lambda\) remains consistently greater than 1, consistent with a scenario in which pressure is applied directly from the loading direction via the culet.

In regions where the NDs, the ruby (pink circle), and the iron sample (black dashed square) are in direct contact with the upper diamond anvil through the ND layer, the red fluorescence contrast is significantly diminished, making ODMR spectra acquisition impossible in those areas.

\subsection{Double-layer PTM Configuration}
\begin{figure}
\centering
\includegraphics[width=\linewidth]{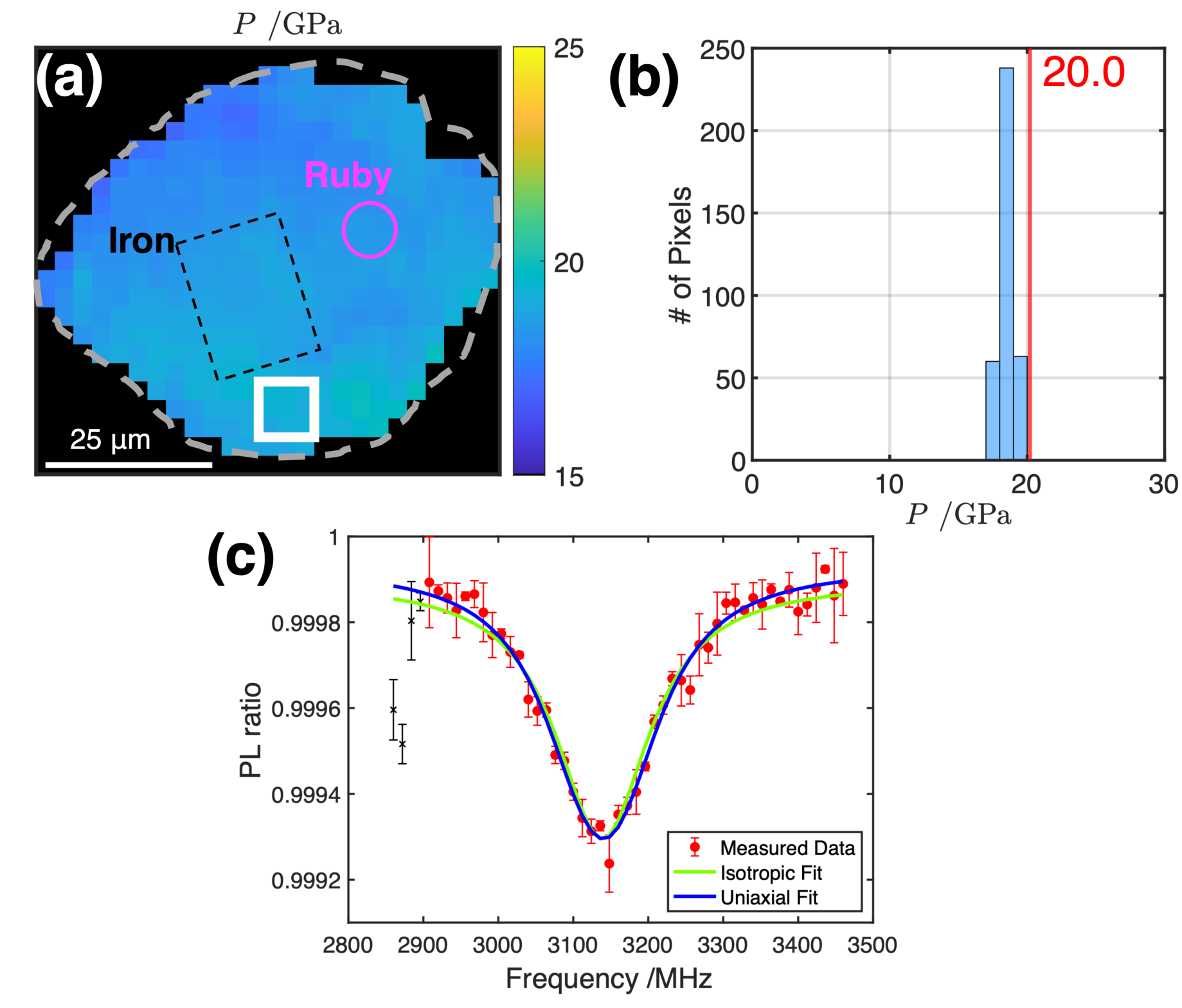}
\caption{\label{fig:4} (Color online) 
Results from the double-layer PTM experiment. 
(a) Spatial distribution of hydrostatic pressure $P$ inside the sample chamber. The gray outline indicates the boundary of the sample chamber. The pink circle and the black box indicate the positions of the ruby and the iron, respectively. Black pixels represent regions with insufficient signal intensity.  
(b) Histogram of hydrostatic pressure $P$ within the sample chamber. The red solid line indicates the pressure value obtained from ruby fluorescence ($20.0\,\mathrm{GPa}$). 
(c) ODMR spectrum acquired from the region marked by the white box in (a).
}
\end{figure}

In contrast to the previous results, the double-layer PTM configuration exhibits a relatively uniform, hydrostatic pressure distribution. Figure~\ref{fig:4}(a) shows the pressure distribution within the DAC sample chamber analyzed by the hydrostatic model. A representative ODMR spectrum is presented in red marks in Fig.~\ref{fig:4}(c). This is acquired at the region marked by the white box in (a). The spectrum has a central frequency of approximately 3150~MHz and a linewidth of about 200~MHz. It displays a symmetric dip shape, which corresponds to a pressure of approximately 18.5~GPa under the hydrostatic model. Signals below 2900~MHz (indicated by black markers), which are again due to the contribution of the NV at ambient pressure, are excluded from the following analysis.

Unlike in the single-layer case, both the hydrostatic model (green curve) and the uniaxial stress model (blue curve) provided similar fitting accuracy. The spectral shape does not exhibit clear asymmetry indicative of uniaxial stress. This behavior is consistently observed throughout the sample chamber. This suggests that the hydrostatic model provides an appropriate description in this case.

The pressure distribution and the resulting histogram obtained from analysis based on the hydrostatic model [Fig.~\ref{fig:S2}] are shown in Figs.~\ref{fig:4}(a) and (b), respectively. The pressure values fall within a relatively narrow range of 17--20~GPa, supporting that a uniform hydrostatic condition is maintained across the whole chamber, including near its boundaries. The red solid line in Figs.~\ref{fig:4}(b) indicates the pressure value of 20.0~GPa obtained from the ruby. The results of the two independent pressure measurements show a satisfactory agreement.

As with the single-layer case, black pixels indicate regions where fitting is not possible due to low signal intensity or insufficient ODMR contrast. In this double-layer case, the ruby and iron samples are not in direct contact with the upper diamond anvil. Consequently, ODMR spectra with sufficient red fluorescence contrast are successfully obtained even in the vicinity of the iron sample. This outcome highlights a marked contrast with the single-layer PTM configuration.

\subsection{Comparison of Two Configurations}
So far we have seen that the number of PTM layers has a significant impact on the degree of hydrostaticity inside the sample chamber. Specifically, a single-layer PTM configuration results in strongly non-hydrostatic conditions, whereas the double-layer configuration maintains a hydrostatic environment.

To conduct a more rigorous comparison, it is instructive to examine what happens when the hydrostatic model is applied to the single-layer case. Despite the fact that the spectra exhibit asymmetry, it is still possible to obtain a pressure distribution by fitting them using a hydrostatic model. In this case, the estimated pressure values are found to lie around 12--14~GPa [Fig.~\ref{fig:S1}]. This result deviates significantly from the pressure measured by ruby fluorescence (20.6~GPa), indicating that the fitting fails to capture the genuine status inside the chamber. 

Conversely, the uniaxial stress model can also be applied to the double-layer case. Although the fitting is less stable due to the increased number of fitting parameters, the extracted anisotropy $\lambda$ ranges from 1.2 to 1.5 [Fig.~\ref{fig:S2}]. This result is clearly different from that of the single-layer configuration (1.5--2.0) shown in Fig.~\ref{fig:3}(f). Therefore, it can be reasonably concluded that the degree of pressure anisotropy is significantly lower in the double-layer configuration than in the single-layer one.

\begin{figure}
\centering
\includegraphics[width=\linewidth]{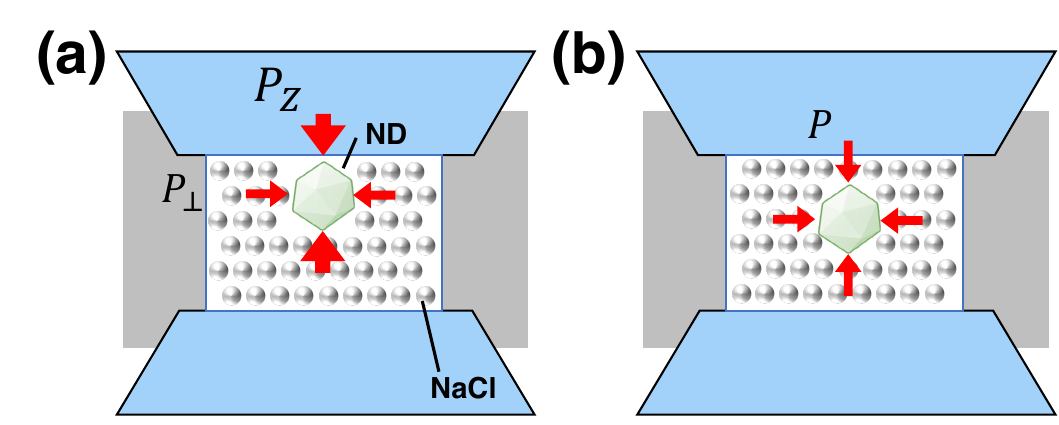}
\caption{\label{fig:5} (Color online) 
(a) Schematic diagram of the sample chamber with a single-layer PTM under uniaxial stress: the NDs are directly compressed by the diamond anvils.
(b) Schematic diagram of the sample chamber with a double-layer PTM under hydrostatic pressure: the additional NaCl layer prevents the NDs from being directly compressed by the anvils.
}
\end{figure}

The fundamental difference between the single- and double-layer configurations is intuitively clear: the top and bottom NaCl layers in the double-layer setup serve to reduce direct contact between the NDs and the diamond anvils [Fig. ~\ref{fig:5}(b)], which is not the case in the single-layer case [Fig. ~\ref{fig:5}(a)]. This configuration likely contributes to the suppression of uniaxial stress on the NDs.

Although such an interpretation may seem self-evident, the ability in this study to visualize local pressure distributions, including the arrangement of the PTM, and to quantitatively evaluate the degree of non-hydrostaticity is particularly important for demonstrating the utility of this method for high pressure research.

\section{Conclusion}
In this study, we demonstrated micrometer-scale imaging of the pressure distribution inside a DAC sample chamber by dispersing NDs and analyzing their ODMR spectra. By applying both hydrostatic and uniaxial stress models, we quantitatively evaluated the influence of the PTM configuration on non-hydrostaticity. These results establish NDs as a high-resolution sensor capable of detecting stress anisotropy, thereby providing a powerful tool for high-pressure physics.

The present approach is not limited to pressure sensing and can be extended to the detection of other physical quantities, such as magnetic fields. We anticipate that this method will contribute to future investigations of material properties under extreme conditions and facilitate the discovery of novel phenomena, including high-temperature superconductivity.

\section*{Acknowledgements}
This work was partially supported by
JST, CREST Grant No.~JPMJCR23I2, Japan; 
Grants-in-Aid for Scientific Research (Nos.~JP25H01248, JP22K03524, JP23K25800, JP24K21194, JP25K00934, JP25KJ1141, and JP25KJ1166); 
Seiko Instruments Advanced Technology Foundation Research Grants; 
Daikin Industry Ltd; 
the Cooperative Research Project of RIEC, Tohoku University;
``Advanced Research Infrastructure for Materials and Nanotechnology in Japan (ARIM)'' (No.~JPMXP1222UT1131) of the Ministry of Education, Culture, Sports, Science and Technology of Japan (MEXT). 
R.S. acknowledges supports from FoPM, WINGS Program, The University of Tokyo.
K.Y. acknowledges supports from JST SPRING Grant (No.~JPMJSP2108), Japan, and MERIT-WINGS, The University of Tokyo.

\nocite{*}
\section*{ } 
\bibliography{refarxiv}
\bibliographystyle{apsrev4-2}

\clearpage
\onecolumngrid
\setcounter{figure}{0}
\renewcommand{\thefigure}{S\arabic{figure}}

\begin{center}
  {\Large\bfseries SUPPLEMENTAL FIGURES AND FURTHER DETAILS}
\end{center}

\begin{center}

  \begin{minipage}{0.48\textwidth}
    \centering
    \includegraphics[width=0.8\linewidth]{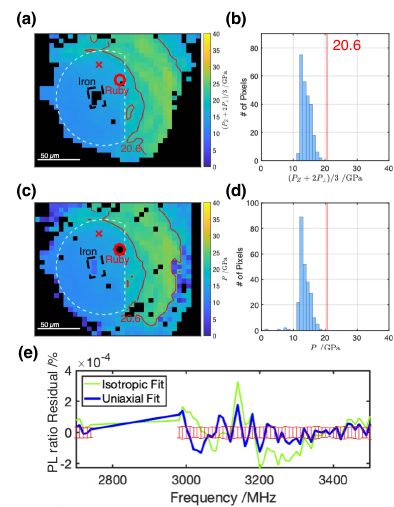}
    \captionof{figure}{Results from the single-layer pressure medium experiment. We analyzed this result using both the hydrostatic model and the uniaxial stress model. 
(a,c) Spatial distribution of pressure [(a)uniaxial stress model:$(P_Z+2P_\perp)/3$ (c)hydrostatic model:$P$] inside the sample chamber. The white outline indicates the sample chamber, the black box marks the iron sample, and the red circle marks the ruby. Black pixels represent regions with insufficient signal intensity.  
(b,d) Histogram of pressure [(b)uniaxial stress model:$(P_Z+2P_\perp)/3$ (d)hydrostatic model:$P$] within the sample chamber. The red solid line indicates the pressure value obtained from ruby fluorescence ($20.6\,\mathrm{GPa}$). (e)Residuals between the experimental ODMR contrast and the two models: hydrostatic (green line) and uniaxial stress (blue line). The red box plots indicate the experimental uncertainty in the ODMR contrast. }
    \label{fig:S1}
  \end{minipage}
  \hfill

  \begin{minipage}{0.48\textwidth}
    \centering
    \includegraphics[width=\linewidth]{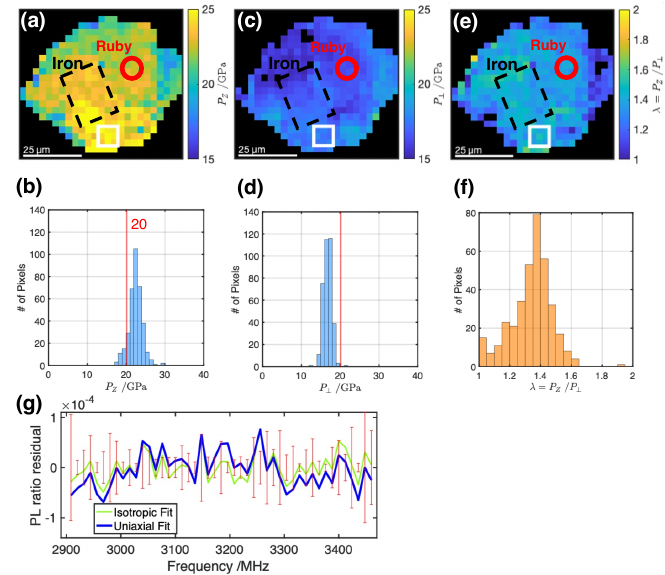}
    \captionof{figure}{Results from the double-layer pressure medium experiment. We analyzed this result using the uniaxial stress model. 
(a, b, c) Spatial distributions of axial pressure $P_Z$, transverse pressure $P_\perp$, and the pressure anisotropy ratio $\lambda = P_Z / P_\perp$. The white dashed line indicates the boundary of the sample chamber, the red circle marks the location of the ruby, the black box indicates the iron sample and black pixels denote regions with insufficient signal intensity. In (a) and (b), the red solid line corresponds to the pressure value of $20\,\mathrm{GPa}$ obtained from the ruby.  
(d) ODMR spectrum at the location marked by a red cross in (a, b). Black points were excluded from the fitting to avoid the influence of ambient pressure components.
(e) Histogram of $P_Z$ and $P_\perp$ within the sample chamber.  
(f) Histogram of the pressure anisotropy ratio $\lambda = P_Z / P_\perp$ within the sample chamber. 
(g) Residuals between the experimental ODMR contrast and the two models: hydrostatic (green line) and uniaxial stress (blue line). The red box plots indicate the experimental uncertainty in the ODMR contrast. }
    \label{fig:S2}
  \end{minipage}
\end{center}
\end{document}